\documentclass{achemso}
\setkeys{acs}{email=false}
\usepackage[english]{babel}
\usepackage[T1]{fontenc}
\usepackage{csquotes}
\usepackage[fleqn]{amsmath}
\usepackage{amsmath}
\usepackage{setspace}
\usepackage{pdfpages}
\usepackage{wasysym}
\usepackage{amsmath}
\usepackage{graphicx}
\usepackage{array}
\graphicspath{figures/}

\title{}

\begin{document}
\singlespacing
\vspace{-1.5in}
\begin{center}
\begin{Large}
\textbf{Quantifying the Economic Case for Electric Semi-Trucks}\\
\end{Large}
\vspace{0.5cm}
\noindent\large{Shashank Sripad\textit{$^{a}$} and Venkatasubramanian Viswanathan\textit{$^{a}$}$^{\ast}$}\\
$^a$Department of Mechanical Engineering, Carnegie Mellon University, Pittsburgh, Pennsylvania, 15213, USA\\
$^\ast$ Corresponding author, Email: venkvis@cmu.edu\\
\end{center}

\begin{abstract}
There has been considerable interest in the electrification of freight transport, particularly heavy-duty trucks to downscale the greenhouse-gas (GHG) emissions from the transportation sector.  However, the economic competitiveness of electric semi-trucks is uncertain as there are substantial additional initial costs associated with the large battery packs required.  In this work, we analyze the trade-off between the initial investment and the operating cost for realistic usage scenarios to compare a fleet of electric semi-trucks with a range of 500 miles with a fleet of diesel trucks.  Here, we define the payback period as the time period required for the operational cost savings from electric semi-trucks to break-even with the initial price differential between the electric and diesel truck. For the baseline case with 30\% of fleet requiring battery pack replacements and a price differential of US\$50,000, we find a payback period of about 3 years or an odometer reading of around 200,000 miles.  Based on sensitivity analysis, we find that the fraction of fleet that requires battery pack replacements is a major factor.  For the case with 100\% replacement fraction, the payback period could be as high as 5-6 years.  We identify the price of electricity as the second most important variable. With an electricity price of US\$0.14/kWh, the payback period could go up to 5 years. Electric semi-trucks are expected to lead to savings due to reduced repairs and magnitude of these savings could play a crucial role in the payback period as well.  With increased penetration of autonomous vehicles, the annual mileage of semi-trucks could substantially increase and this heavily sways in favor of electric semi-trucks, bringing down the payback period to around 2 years at an annual mileage of 120,000 miles.  There is an undeniable economic case for electric semi-trucks and developing battery packs with longer cycle life and higher specific energy would make this case even stronger.
\end{abstract}


\section*{Introduction}
\vspace{-10pt}
There is an enormous interest around the electrification of Class 8 heavy-duty trucks following the unveiling of the Tesla Semi\cite{tesla} and announcements by several other auto-and-truck manufacturers including Cummins, Daimler, BYD and Volvo.\cite{cummins,daimler,byd,volvo} The importance of the trucking industry is highlighted by its share of the total freight shipments which is about 65\% by value and 68\% by weight \cite{birky2017transportation} within the United States. At the same time, 24\% to the greenhouse-gas (GHG) emissions from the transportation sector\cite{davis2017transportation} is due to the trucking industry. Batteries play a crucial role in enabling the transition to electric transportation which can mitigate GHG emissions and potentially reduce the energy consumption.\cite{crabtree2015perspective}  In an earlier work, we analyzed the performance requirements of Li-ion batteries for electrifying class 8 semi-trucks.  We demonstrated that semi-trucks would be limited to a driving range of under 600 miles in order to be able to carry an average payload of about 16 US tons\cite{sripad2017performance} and we highlighted the significantly higher costs incurred due to the large battery packs required.  In a follow-up work, we quantified the potential for platooning to reduce the performance requirements of batteries and estimated a 15\% reduction in the energy requirements.\cite{guttenberg2017evaluating}  Both of these works were widely covered in the media and were largely consistent with the estimates that Tesla unveiled.\cite{wired,wapo,reuters2,mittr}  A final missing piece is to quantify and compare the total and operational costs of electric and diesel semi-trucks in different realistic scenarios and explore the economic case for electric semi-trucks.

Class 8 semi-trucks have a typical lifetime mileage of around 1 million miles.\cite{davis2017transportation,atri2017,nrcreport}  The average annual miles for the first two years of operation is over 100,000 miles \cite{davis2017transportation} which decreases with the age of the truck and the average annual mileage for the semi-truck fleet in the U.S is about 75,000 miles\cite{nrcreport}.  Studies have shown that for a representative sample set of semi-trucks, about $\sim$40\% are known to travel well over 500 miles and an equal fraction travel between 100-500 miles per trip.\cite{atri2017,nrcreport} Hence, in order to perform a realistic comparison, we choose a 500-mile capable electric semi-truck and a typical diesel semi-truck with a range of about 1000 miles. 

The main cost categories for class 8 semi-trucks are the initial investment, operational costs and the periodic costs that occur due to major replacements.  A major benefit for electric trucks is the significantly lower operational costs due to two factors: (i) increased energy efficiency for mobility and (ii) similar or lower cost per unit energy for electricity compared to diesel.  Undoubtedly, there could be economic consequences to the potential reduction in payload carrying capacity due to the additional weight of the battery pack relative to diesel as a fuel.  There are two scenarios are commonly referred to as `cubing-out' and `grossing-out', where cubing-out refers to exceeding the volume limit of the trailer while grossing-out refers to reaching the gross-weight limit.  The reduced payload carrying capacity of electric trucks only applies to the grossing-out scenario.  It is worth highlighting that many fleets cube out or have limited cargo per payload and as a result, don't gross out.\cite{davis2017transportation}  However, in this analysis, we ignore the potential economic loss associated with trucks grossing-out and plan to revisit this in future studies.

The central question that we address in this work is the trade-off between the initial investment cost and the operating cost for realistic usage scenarios and identify the payback period for an electric semi-truck.  Further, we analyze the economic benefits of platooning and how different variables affect the payback period. It is extremely important to understand and quantify the payback period in order to inform and facilitate mainstream adoption of electric heavy-duty vehicles.

\begin{figure}[t]
 \centering
 \includegraphics[width=\linewidth]{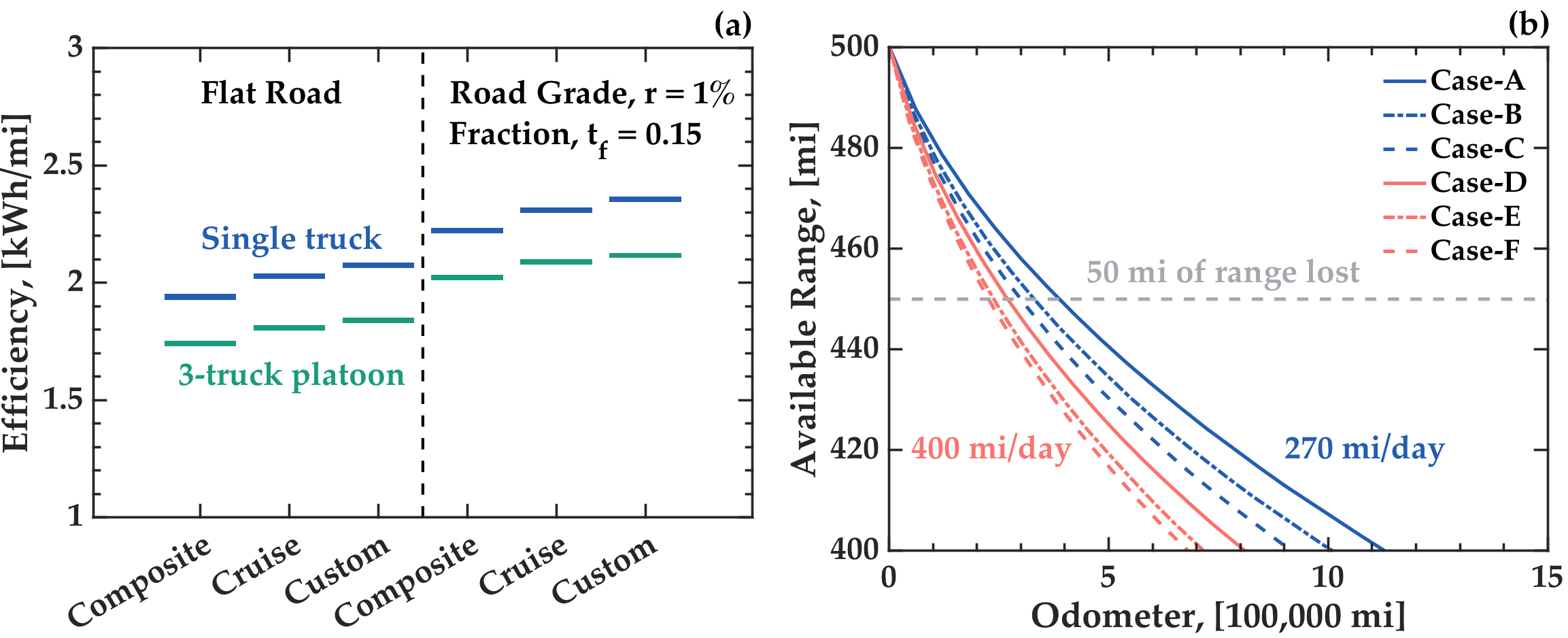}
 \caption{\textbf{The energy consumption characteristics under different road conditions for a single truck with a drag coefficient of 0.4 and a platoon of 3 trucks is shown in Part (a) for a road grade, (r), over a specified fraction of the total distance, (t$\mathrm{_f}$). We estimate the energy consumption to be 2.05$\pm$0.32 kWh/mi over all the use cases considered. The drive cycles are shown in Figure (S2).  Part (b) shows the estimates for the cycle life with all Cases running under full load conditions. Case A and D represent 3 truck platoons with the Composite duty-cycle under flat road conditions with regular charging, representing the optimistic scenario. Case B and E refer to single trucks with the Cruise duty-cycle also under flat road conditions with faster charging rates. Case C and F refer to single trucks with the Custom duty-cycle at 1\% road grade for a small fraction of the trip along with fast charging infrastructure representing pessimistic conditions. As the total mileage increases, the available range decreases monotonically as a function of the duty cycle the battery pack is subjected to. The optimistic usecases, with platooning\cite{crabtree2017transportation} and regular charging show a much higher cycle life than single trucks with fast charging. Considering all the cases, Part (b) shows that there is a high likelihood of battery packs reaching end-of-life before the lifetime of the truck.}}
\label{figure1}
\end{figure}

\section*{Performance of Electric and Diesel Semi-Trucks}
\vspace{-10pt}
In an earlier work\cite{sripad2017performance}, we used a parameterized vehicle dynamics model to quantify the energy requirements of a class 8 semi-truck. Details of vehicle design as well as performance parameters of the battery pack form the primary inputs to such a model described in \textit{Methods}.  Accounting for possible highly optimized vehicle designs, in this work we utilize a similar model with a drag coefficient of 0.40$\pm$0.04, rolling resistance  0.0075$\pm$0.002 and other parameters representative of a Class 8 semi-truck.\cite{icct_supert,sripad2017performance} Using these parameters, along with realistic drive cycles\cite{nrel_dc} shown in Figure (S2), we estimate that a well-designed electric semi-truck under full-load conditions would have an energy consumption of 2.05$\pm$0.32 kWh/mi or 51.25$\pm$8 Wh/ton-mile depending on the road conditions and the duty cycle, as shown in Figure (\ref{figure1}). On the other hand, for conventional diesel trucks, fleets achieve an efficiency in the range of [6-8.5] mpg,\cite{atri2017,nrcreport,davis2017transportation,nptcreport} and considering the energy content in diesel, this turns out to be about [4.45-6.3] kWh/mi.

Based on the results shown in Figure (\ref{figure1}a), the battery pack required for a range of 500 miles is about $\sim$1,000 kWh. Current estimates of the price of battery packs are about US\$[180-220]/kWh\cite{bnef,kittner2017energy,sakti2017consistency} while the cells of certain chemistries have been reported to have reached prices below US\$100/kWh.\cite{bnef} Some forecasts for the price of battery packs by the year 2020 are around US\$125/kWh.\cite{kittner2017energy} In the 2020 time-frame, battery packs for 500-mile electric trucks would cost about US\$125,000 or lower. For a longer time-frame, we assume a price in the range of US\$[90-120]/kWh for battery packs.\cite{bnef}

For the case of electric semi-trucks, assuming 260 driving days in a year, based on the limits of the annual mileage, the daily distance traveled is in the range of 270 and 400 miles. Considering the daily distance driven, along with the different drive cycles derived from CARB-HHDDT\cite{nrel_dc} and road conditions, we obtain different duty-cycles that the battery pack is subjected to. The various realistic duty-cycles are used to estimate the cycle life of the battery pack. Figure (\ref{figure1}b) shows the cycle life estimates from the battery pack simulations. The simulations are performed on a full-pack model which consists of cells based on a high specific energy chemistry, NMC-622 ($\mathrm{LiNi_{0.6}Mn_{0.2}Co_{0.2}O_2}$) cathode and Graphite anode. The details of the battery pack and degradation model along with the specification of the cells can be found in \textit{Methods} and \textit{Supplementary Information}. As seen in Figure (\ref{figure1}b) the available range for all the cases apart from Case-A and Case-B reduces to under 400 miles before the total distance traveled reaches 1 million miles. An available range of 400 miles also represents 80\% of the initial capacity, generally considered the end-of-life criterion\cite{peterson2010lithium}. The use of fast charging, i.e. charging rates of over 2C within a CC-CV protocol is seen to reduce the cycle life significantly. Platooning, represented by Case A and D where the energy consumption is reduced when coupled with lower charging rates is seen to improve the cycle life corroborating the results from our previous work\cite{guttenberg2017evaluating} where a different example semi-truck was studied. Figure (\ref{figure1}b) shows that likelihood of the battery pack needing a replacement is high for all cases considered.  Hence, we believe that the fraction of the fleet requiring a battery pack replacement is an important variable.  

The end-of-life condition for electric semi-trucks requires a careful consideration as trucks could be reassigned to routes according to their available range.  Hence, although, nearly all cases considered yield an available range of 400 miles well below the 1 million mile mark, it is likely that re-balancing the routes among the fleet could ensure that not all of the fleet vehicles need a battery replacement.  Given the importance of the requirement for battery replacement, we carry out a rigorous sensitivity analysis with respect to battery replacement that we will discuss later.

\section*{Economics of Electric and Diesel Semi-Trucks}
\vspace{-10pt}
In terms of economic variables, the general operational costs (General Op\textendash Costs), common to both electric and diesel trucks are the cost of the cab, driver wages, insurance, tire replacements, and the cost of the permits and tolls, totally amounting to US\$[0.76-0.81]/mi\cite{atri2017,nptcreport}. Electric powertrains are expected to have much lower repairs compared to diesel powertrains which comprise of several moving parts along with the engine which require frequent maintenance.  The additional repairs that diesel trucks are considered in the range of US\$[0.15-0.16]/mi.\cite{atri2017} The nominal price of diesel is considered with known projections and is seen to be in the range of US\$[2-4]/ga which is about US\$[0.05-0.11]/kWh of diesel.\cite{eiadiesel,eia2018aeo} A discount rate of 3\% is used throughout the cost model. The price of electricity is considered in the range of US\$[0.07-0.12]/kWh.\cite{eia2018electric,eia2018aeo} Each of the above-mentioned variables with their stipulated bounds form the baseline scenario for analyzing the total ownership costs.

\begin{figure}[t]
 \centering
 \includegraphics[height=7cm]{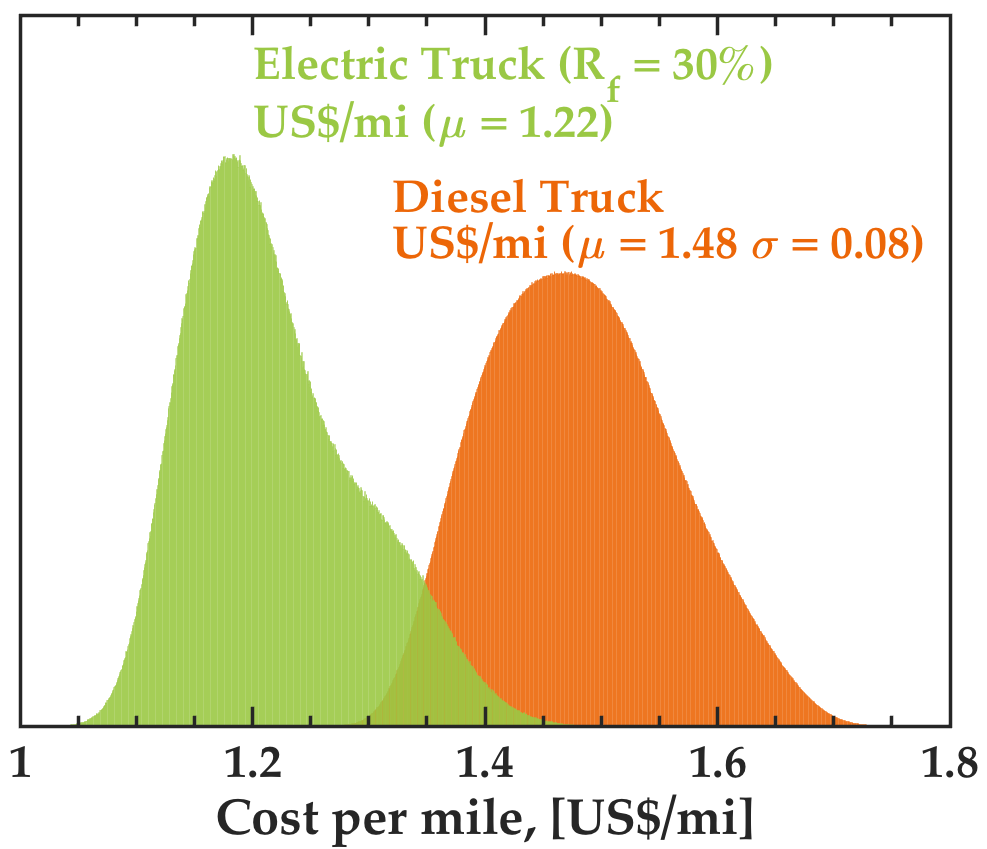}
 \caption{\textbf{Comparison of the total cost of ownership per mile for a diesel truck fleet and an electric truck fleet with a battery replacement fraction, ($\mathrm{R_f}$), of 30\%. The cost per mile for $\mathrm{R_f}$=0\% is US\$1.18$\pm$0.05/mile, and at $\mathrm{R_f}$=100\% is increases to US\$1.3$\pm$0.05/mile. For the baseline scenario, described in Table (\ref{tbl}), the cost per mile for the diesel truck fleet is about US\$1.48$\pm$0.08/mile and that of the electric truck fleet with $\mathrm{R_f}$=30\% is about US\$1.22/mile. We observe a small region of overlap of the distributions for the most favorable cases for the diesel trucks and the most unfavorable cases for the electric trucks.}}
 \label{figure2}
\end{figure}

The total operational costs of electric and diesel semi-trucks is shown in Figure (\ref{figure2}). All variables are uniformly discretized within the bounds where each discrete value is assumed to have an equal probability of occurrence, thereby providing the distributions in Figure (\ref{figure2}). The cost per mile for diesel trucks is US\$1.48$\pm$0.08/mile and that of the electric truck is determined to be US\$1.22/mile under the baseline scenario of with a battery replacement fraction, ($\mathrm{R_f}$), of 30\%. The price of the battery pack is assumed to be in the range of US\$[90-120]/kWh, as discussed previously. In the most favorable scenario for electric trucks, with low electricity prices, high efficiency due to flat roads and/ or platooning, the cost per mile could be as low as US\$1/mile while a correspondingly favorable scenario for diesel trucks results in a cost of US\$1.3/mile. The distribution for electric trucks in Figure (\ref{figure2}), is skewed due to the cases with battery pack replacements. As the pack replacement fraction increases the mean cost per mile increases linearly to about US\$1.3/mile at the limiting case $\mathrm{R_f}$=100\%. This is visualized in Figure (S4,S5).

\section*{Payback Period:}
\begin{figure}[t]
 \centering
 \includegraphics[height=7cm]{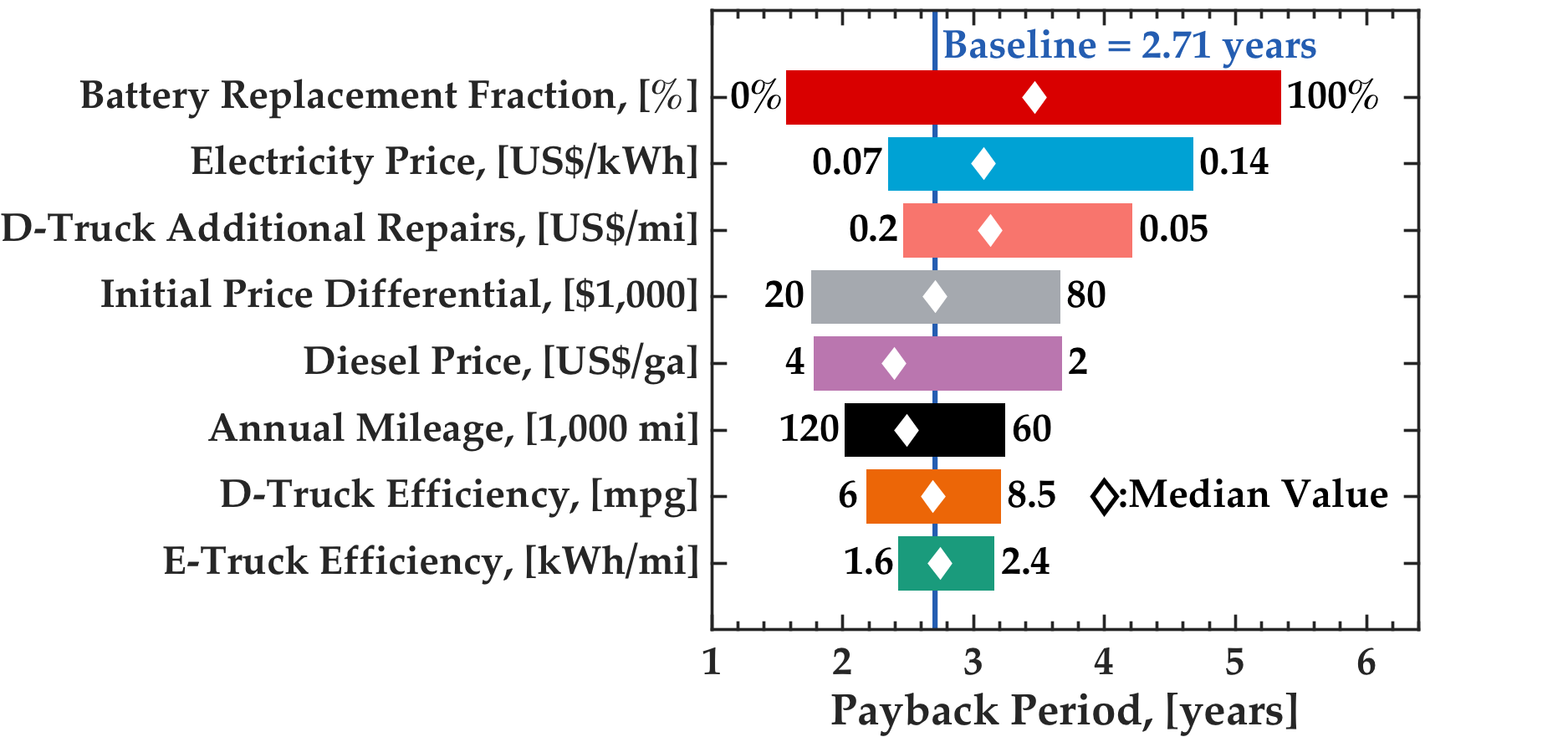}
 \caption{\textbf{Quantifying the sensitivity of the payback period to different variables. The variable under consideration is pinned to a constant value while the other variables remain at baseline values compiled in Table (\ref{tbl}). The mean value of the payback period distribution in the baseline scenario, of 2.71 years is shown above. The sensitivity of the payback period to variables like the battery replacement fraction is large, primarily due to price and size of the battery packs that need to be replaced. The price of electricity has a significant impact on the payback period and high electricity prices could increase the payback period by about 75\%. The sensitivity around E-Truck efficiency includes the case of platooning where low rates of energy consumption like 1.6 kWh/mi can be achieved, although the payback period reduces by only 0.3 years.}}
 \label{figure3}
\end{figure}

\vspace{-10pt}
An important metric that needs to be studied is the payback period. In the context of this study, we define the payback period as the time required for the cost savings from the operation of electric semi-trucks to break-even with the initial price differential between the electric and diesel truck.  The payback period distribution obtained for the baseline scenario exhibits a mean value of 2.71$\pm$0.7 years. The sensitivity of the payback period to several salient variables can be seen in Figure (\ref{figure3}), where we can analyze the payback period both at the bounds of variables as well as the median value of variable range considered. The distribution for the payback period in the baseline scenario can be found in Figure (S3).


\begin{table}
\small
 \caption{\ \textbf{Values of variables that define the baseline scenario to compute the total operational costs and payback period. General Op\textendash Costs do not influence the payback period since they are equal for both diesel and electric trucks.}}
 \label{tbl}
\small
\centering
\setlength{\tabcolsep}{2mm}
\begin{tabular}{|p{150pt}|p{120pt}|}
\hline
    D-Truck, Initial Price & US\$150,000\cite{posada2016costs}\\
    E-Truck, Initial Price & US\$200,000\\
    Diesel Price & US\$[2.21-4.19]/ga\cite{eiadiesel,eia2018aeo}\\
    Electricity Price & US\$[0.07-0.12]/kWh\cite{eia2018electric,eia2018aeo}\\
    E-Truck, Efficiency & [1.7-2.3]kWh/mi\\
    D-Truck, Efficiency & [6-8.5]mpg\cite{atri2017,nrcreport,davis2017transportation,nptcreport}\\
    D-Truck, Additional Repairs & US\$[0.15-0.16]/mi\cite{atri2017}\\
    Annual Mileage & [80,000-100,000]mi\cite{davis2017transportation,nrcreport}\\
    General Op\textendash Costs & US\$[0.76-0.81]/mi\cite{atri2017,nptcreport}\\
    Battery Pack Price & US\$[90,120]/kWh\cite{kittner2017energy,bnef}\\
    E-Truck, Battery Replacement & 30\% of the fleet\\
    Discount Rate & 3\%\\
\hline
\end{tabular}
\end{table}

As seen in Figure (\ref{figure3}), due to the size and price of battery packs, the fraction of cases that require battery pack replacements shows the highest sensitivity. The limiting scenarios of no replacements and all replacements show payback periods of 1.57 and 5.25 years respectively. A replacement fraction of about 50\% results in a payback period of $\sim$3.5 years. In terms of the price of fuel (diesel or electricity), we observe that increasing the price of electricity from the baseline scenario to about US\$0.14/kWh nearly doubles the payback period. It is worth highlighting that the price of electricity in several locations within the United States is well-above US\$0.14/kWh \cite{eia2018electric,eia2018aeo} and the large sensitivity exhibited is an important factor to consider for the charging infrastructure. This could effectively limit the locations where electric trucks can be utilized. Very low diesel prices would increase the payback period, although, the lower limit of US\$2/ga considered in Figure (\ref{figure3}) is much lower than current projections and estimates.\cite{eiadiesel,eia2018aeo} Further, it could be concluded that the economic case for electric trucks is influenced to a much larger extent be the price of electricity than by the price of diesel, given the known projections for the possible fluctuations in the prices. Both the price of electricity and that of diesel show a significant impact on the payback period.  A brief discussion on the comparison between electricity and diesel in terms of the price of per unit energy is compiled in the \textit{Supplementary Information} and additional information is shown in Figure (S1).

Among other variables seen in Figure (\ref{figure3}), the additional repairs required by diesel trucks, which form less than 10\% of the cost per mile of diesel trucks, is seen to affect the payback period by over one year.  The sensitivity of the payback period to the initial price differential shows a clear linear relationship and at a large price differential of US\$80,000, we see that the payback period increases to about 3.8 years. The annual distance traveled also influences the payback period to a significant extent. If the annual mileage is under 100,000 miles, the payback period would be extended, with a payback period of about 3.2 years at an annual mileage of 60,000 miles. On examining the payback period for the median value, we observe a non-linear relationship between the payback period and the annual mileage. In terms of energy efficiency, for the same range of conditions considered, the potential variation in diesel truck efficiency affects the payback period to a larger extent than that of its electric counterpart. The electric efficiency variables considered include the scenario of platooning to reach energy consumption rates as low as 1.6 kWh/mi. While the higher efficiency of electric trucks results in a much lower cost per mile over the lifetime of the truck, the payback period does not show a significant sensitivity towards the same.

It is worth noting that there are several other variables like changes in wages due to automation, tires with reduced wear-and-tear, reduced insurance costs etc., which affect the operational costs of electric and diesel trucks in a similar manner and hence, such variables do not influence the payback period. In addition, factors such as the charging infrastructure\cite{limb2016economic} could add significant costs to the electric trucks, albeit, it could be argued that similar costs are not accounted for with the diesel trucks in terms of gas/ diesel stations. Extending this argument further, it might be imperative for OEMs or external entities to be responsible for the costs of the charging infrastructure in order to ensure an even-handed and favorable economic case for the electrification of semi-trucks.

Throughout this work, we consider vehicle designs with a low drag coefficient of around 0.4, and it should be noted that most semi-truck designs currently in the market have drag coefficients of well-over 0.5 \cite{icct_supert,nrcreport}. Enabling a range of about 500 miles for vehicle designs with high drag coefficients without sacrificing the payload capacity is possible only with very high specific energy battery packs.\cite{sripad2017performance} Higher drag coefficients result in high energy consumption requiring larger battery packs which effectively increase the initial price differential. A vignette on the trade-offs between vehicle design and specific energy for a fixed range and their effect on the payback period is compiled in the \textit{Supplementary Information} and Figure (S3). If the drag coefficient is about 0.63, which is the current fleet average for the U.S.,\cite{nrcreport} the mean payback period increases to about 8.45 years which is close to the lifetime of the truck itself thereby suggesting that there exists a vehicle design constraint for commercial viability.

\section*{Is there an Economic Case for Electric Trucks?}
\vspace{-10pt}
Across all the variables considered, the payback period is most sensitive to the battery pack replacement fraction which is due to the size and effectively the price of battery packs. This implies that for high-utilization applications like semi-trucks where the battery packs are extensively used, higher cycle life coupled with a lower purchase price of battery packs will play a crucial role in creating a favorable economic case. The next variable of importance is the price of electricity where the payback period could increase by about 2 years from the baseline scenario if the price of electricity reaches US\$0.14/kWh. The variable that represents additional repairs for diesel trucks also has a significant impact on the payback period. If the repairs for diesel trucks are only US\$0.05/mi higher than that of electric trucks, the payback period could increase by $\sim$1.5 years from the baseline to $\sim$4.2 years. Apart from the above-mentioned variables, with the rest of variables, even the unfavorable scenarios result in a payback period of under 4 years. On the other hand several favorable scenarios could result in a payback period of 2 years or lower.\\

In the overall analysis, it is evident that there exists a strong economic case for the adoption of electric heavy-duty commercial vehicles particularly for driving ranges of up to 500 miles. In the premise for this work, we highlighted the importance of the trucking industry and also its considerable contribution to the greenhouse gas emissions. While it is crucial to accelerate the adoption of electric semi-trucks, it should also be noted that there are important factors like the price of the battery pack, their cycle life and the price of electricity within the charging infrastructure. Each of these variables should be carefully considered in order to ensure favorable economics. Also, a well-designed exterior with low energy consumption and high specific energy battery packs will be key in enabling electric semi-trucks with driving ranges of up to 500 miles. Increasing the driving range further without affecting the commercial viability would require higher specific energies and lower battery pack prices while other important factors mentioned previously remain favorable.

\clearpage
\section*{Abbreviations}
\vspace{-10pt}
CARB-HHDDT: California Air Resources Board-Heavy Heavy Duty Diesel Truck\\
D-Truck: Diesel Truck\\
E-Truck: Electric Truck (Range of 500 miles)\\
Op\textendash Costs: Operational Costs\\
CC-CV: Constant Current-Constant Voltage\\
OEM: Original Equipment Manufacturer\\
mi: mile(s)\\
ga: gallon(s)

\section*{Methods} 

\begin{footnotesize}
\textbf{Vehicle Dynamics:} The power demands of the electric semi-truck are estimated using a parametric relationship between the vehicle design parameters, the road conditions and the drive cycle in consideration, as shown by: 
\vspace{-0.35cm}
\begin{gather*}
\;\;\;\;\;\;\;\;\;\;\;\;\;\;\;\;\;\mathrm{P(t) = \bigg(\frac{1}{2}\rho.C_{d}.A.v(t)^{3} 
+ C_{rr}.W_{T}.g.v(t)
+ t_{f}.W_{T}.g.v(t).Z
+ W_{T}.v(t).\frac{\textit{d}v}{\textit{d}t}
\bigg)\dfrac{1}{\eta_{bw}}} \;\;,\\
\;\;\;\;\;\;\;\;\;\;\;\;\;\;\;\;\;\mathrm{P_{reg}(t) = \bigg(W_{T}.v(t).\frac{\textit{d}v}{\textit{d}t}\bigg)\eta_{bw}.\eta_{brk}} \;\;,
\label{eq-power}
\end{gather*}
The drive cycle provides the profile of the instantaneous speed, (v(t)). The drive cycles used in this study are shown in Figure (S2). The vehicle design parameters like frontal area, (A), coefficient of rolling resistance, ($\mathrm{C_{rr}}$), are obtain from current data on the fleet of Class 8 trucks in the U.S.\cite{icct_supert,nrcreport} The road gradient, (Z), and the fraction of the trip for which positive road gradients exist, ($\mathrm{t_f}$), are fixed according to the case in consideration. The total weight of the truck, ($\mathrm{W_T}$), is fixed at 80,000 lbs ($\sim$36,360 kg). The other variables include the battery-to-wheels efficiency, ($\mathrm{\eta_{bw}}$), and the efficiency of the brakes, ($\mathrm{\eta_{bw}}$). The regenerative power, ($\mathrm{P_{reg}(t)}$), is used for segments of deceleration and charge rates at the regenerative segments is limited to 2C. The power load obtained for the given case for a stipulated distance is then applied on the battery pack model to study the pack performance.

\noindent \textbf{Battery Modeling:} The battery pack is modeled within AutoLion-ST\cite{kalupson2013autolion} which uses a thermally coupled battery model for each cell. The mathematical relationships and the modeling framework can be found elsewhere.\cite{alst1,alst2,alst3} The cells are assembled into the battery pack but no cell-to-cell variation is considered within the model. The degradation model which accounts for the loss of Li-ions over cycling due to various parasitic processes is implemented as a sub-model within the battery pack model\cite{wang_degradation}. The rate constants of the degradation reactions/ processes are fit to a specific cell chemistry based on NMC-622 ($\mathrm{LiNi_{0.6}Mn_{0.2}Co_{0.2}O_2}$) cathode and Graphite anode which is representative of the high specific energy cells that are required for the electrification of heavy-duty vehicles. Additional details on the degradation model can be found in the \textit{Supplementary Information}.

\noindent \textbf{Cost Modeling:} Total operational costs including the fuel costs are calculated over the total distance traveled for each discrete value of the variables within the bounds from the baseline scenario. All the operational costs are expressed per mile and using the values and bounds of the annual mileage, the annual cash flow distribution is obtained. The present value factors are calculated using a fixed discount rate. Applying the present value factors on the fixed initial investments and the annual cash flows, we obtain the levelized annual costs. Finally, the cost per mile distribution is obtained based on the annual mileage for the respective cases.

The timeline for fixed costs related to battery pack replacements are estimated using the results of the cycle life simulations. The corresponding cash-flows for the battery replacement are converted to present value using the same discount rate. The fraction of cases that require replacement are randomly sampled from the cost per mile distribution to obtain the cost per mile distribution for a fixed replacement fraction. The effect of the battery pack replacement fraction on the cost per mile distribution is captured in the Figure (S4,S5).

The payback period which is the time period over which the fuel savings from the electric semi-truck is able to recover the initial price differential is studied using an approach similar to the cost per mile calculations. The sensitivity analysis for the payback is performed by holding the variable in consideration fixed while the rest of the variables remain at baseline scenario values.
\end{footnotesize}

\section*{Acknowledgements}
\vspace{-10pt}
The authors would like to thank Michael Cembalest and Zach Long at JP Morgan for the extremely helpful discussions about the assumptions in the cost model, Jake Christensen for general critique and insights on the work and Joshua Switkes for valuable comments on platooning and the performance of semi-trucks.

S.S. and V.V. gratefully acknowledge support from the Technologies for Safe and Efficient Transportation University Transportation Center. V.V. gratefully acknowledges support from the Pennsylvania Infrastructure Technology Alliance, a partnership of Carnegie Mellon, Lehigh University, and the Commonwealth of Pennsylvania's Department of Community and Economic Development (DCED).

\bibliography{references} 

\clearpage


\section*{Supplementary material}

# Supplementary Information:
# Quantifying the Economic Case for Electric Semi-Trucks

Shashank Sripad[a] and Venkatasubramanian Viswanathan[a*]
[a]Department of Mechanical Engineering, Carnegie Mellon University,
Pittsburgh, Pennsylvania, 15213, USA
* Corresponding author, Email: venkvis@cmu.edu

## Price of Electricity and Diesel:

One of the major advantages of electric powertrains is the higher efficiency. If we examine the cost of energy contained in the fuel itself, based on the energy density of diesel which translates to about 37.7 kWh/ga. Comparing the price of electricity and diesel using a common scale, we arrive at the comparison shown in Figure (S1).

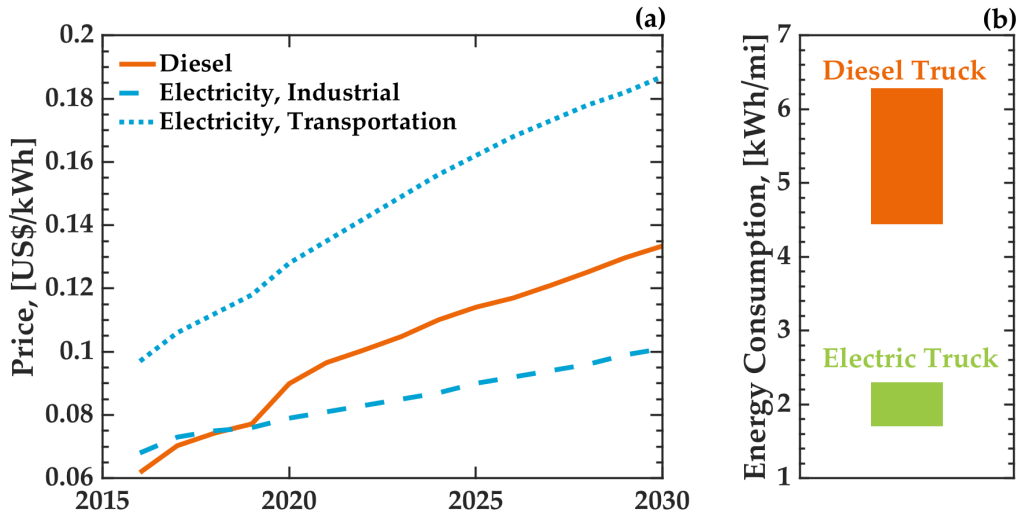

Figure S1: A comparison of the nominal price of fuel per unit energy of diesel and electricity (transportation and industrial) with known projections[1] is shown in Part (a). While the price per unit energy is very similar, the efficiency of the electric powertrain is several times higher than one powered by diesel. As shown in Part (b), for the baseline scenario consideration, if the price per unit energy of electricity and diesel and equal, it is about **2.5-4 times more expensive to power a diesel truck than a well-designed electric truck**.

# Realistic Usecases and Drive Cycles:

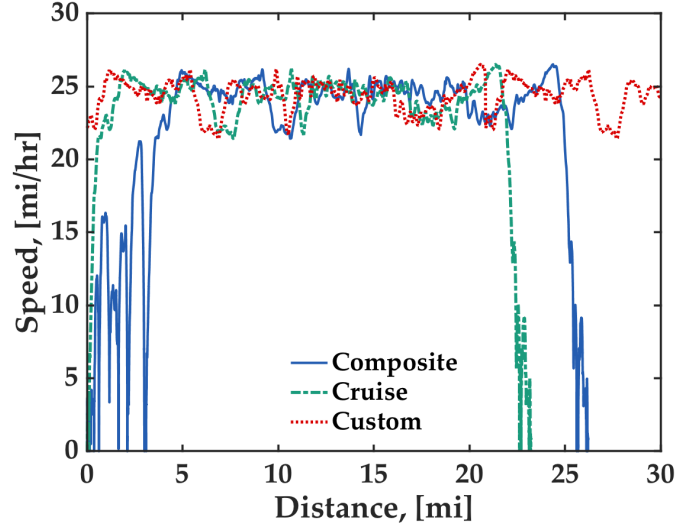

**Figure S2:** The Composite, Cruise and Custom drive cycles used to study the performance of the electric semi-truck is shown over a small representative distance. The drive cycles are repeated over the total trip distance. The Composite and the Cruise drive cycles are segments of the CARB-HHDDT from NREL DriveCAT[2] and the Custom drive cycle is based on the Cruise drive cycle without the acceleration from stop and deceleration to stop segments. The Custom drive cycle is representative of long-haul duty cycles where speed remains close to a mean value for most of the trip.

# Battery Modeling and Simulation:

The degradation processes are modeled within the battery pack model[3–6] degradation sub-model[6] shown below:

$$j_{SEI} = -k_{o,SEI} \cdot c^s_{solvent} \cdot \exp\Big[-\frac{\alpha_{c,SEI} \cdot F}{RT} \cdot \big(\phi_s - \phi_e - I \cdot R_{film} - U_{SEI}\big)\Big]$$

$$j_{PL} = -i_{o,PL} \cdot \exp\Big[-\frac{\alpha_{c,PL} \cdot F}{RT} \cdot \big(\phi_s - \phi_e - I \cdot R_{film}\big)\Big]$$

$$\frac{d\epsilon_{AM}}{dt} = -k_{AMI} \cdot I_{total}$$

where the side currents for each of the degradation processes for Solid-Electrolyte Interphase (SEI), ($j_{SEI}$), for the Lithium plating, ($j_{PL}$), and the last rate equation captures the Active Material Isolation along with the total current, (I). The other constants from the degradation sub-model are the rate constants ($k_{o,SEI} = 1 \times 10^{-12}$ m/s)[6], ($k_{AMI} = 2 \times 10^{-14}$ m/s) and the exchange current density, ($i_{o,PL} = 0.001 A/m^2$)[6]. The ($\alpha$'s) are the cathodic transfer coefficients. ($c^s_{solvent}$), is the concentration of the solvent. The ($\phi$'s) are the potentials of the electrode and liquid phases. ($R_{film}$) is the resistance of the SEI layer.

# Vehicle Design Considerations:

The price of fuel (electricity and diesel) coupled with the efficiency of electric powertrains would always result in lower operational costs for electric trucks. However, an aerodynamically inefficient truck electric truck design would result in a higher energy consumption and require a larger battery pack for a fixed range.[7] A larger required battery pack results in a higher initial price differential and the marginally higher operational costs together result in a much higher payback period distribution, as shown in Figure (S3). A mean payback period of 8.45 years which is very close to the lifetime of the truck itself. Also, it is worth noting that at a drag coefficient of 0.63, the battery pack would be extremely heavy and resulting in reduced payload capacity unless very high specific energy battery packs.

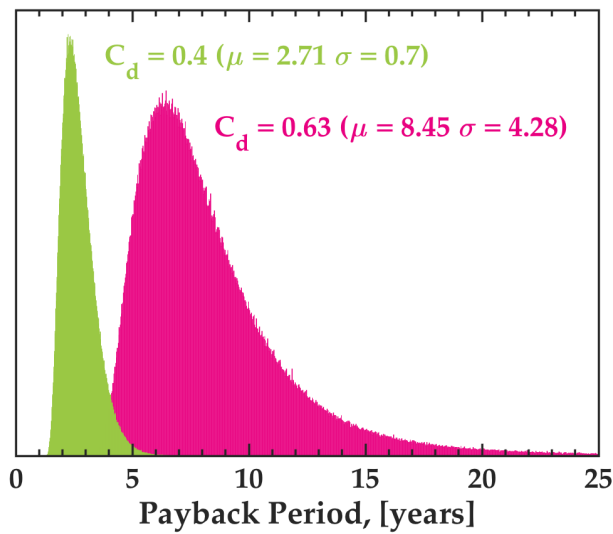

**Figure S3:** The exterior design and the effective drag coefficient of the electric semi-truck plays a crucial role in the economic case, where a lower drag coefficient which results in a lower energy consumption and effectively a smaller battery pack which in-turn decreases the initial price differential. We identify a drag coefficient, $C_d$, of 0.63 to represent the threshold beyond which there is a very high probability of the payback period being higher than the lifetime of the trucks. Although, the case presented above stands for an electric truck with a range of 500 miles or over and a battery pack price in the range of $[90,120]/kWh.

# Battery Pack Replacement:

In order to account for the battery pack replacements, the fraction of cases that require replacement is randomly sampled from the ideal distribution with no replacement and replaced with another random sample with the same number of cases from the limiting distribution with all battery packs replaced. This process is captured in Figure (S4).

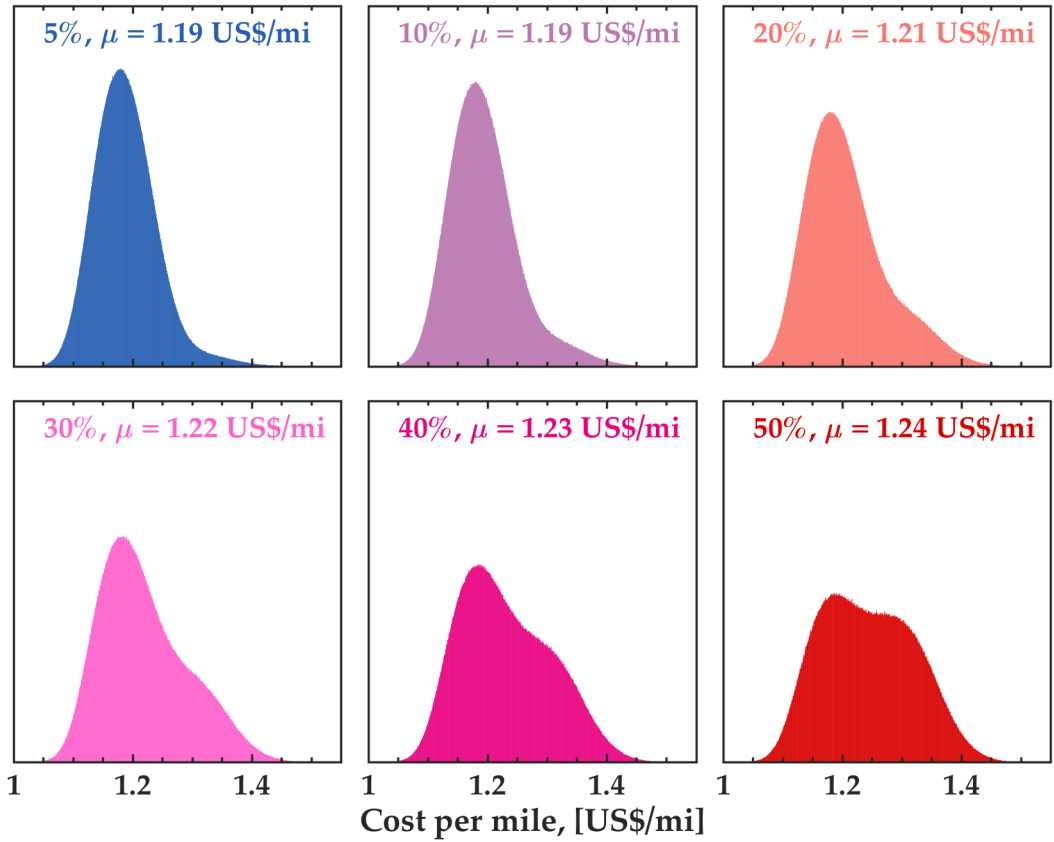

Figure S4: The extent to which the distributions are skewed by the replacement fraction is shown above. The mean value shows a steady increase with the increase in the battery pack replacement fraction as the distributions tend to become bi-modal in nature.

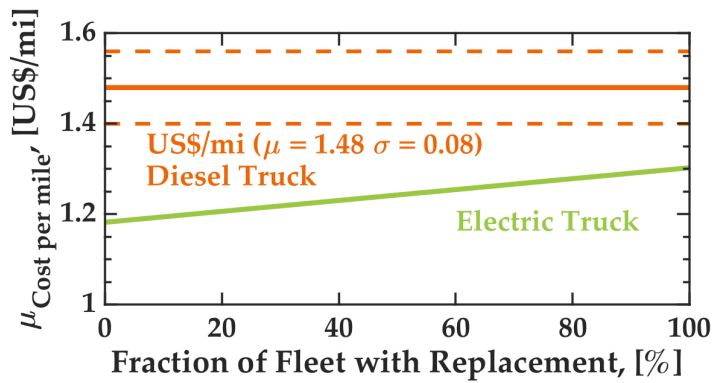

Figure S5: A comparison between the cost per mile of an electric semi-truck fleet and that of a diesel truck fleet is shown as a function of the battery pack replacement fraction. It can be seen that even under the limiting case of all trucks requiring a battery pack replacement, the mean cost per mile is less than the cost per mile of a diesel truck fleet.

4



\end{document}